\documentclass[aip,pop,twocolumn,superscriptaddress]{revtex4}
\usepackage{graphics}
\usepackage{epstopdf}
\usepackage{epsfig}
\usepackage{amsmath}
\usepackage{amssymb}
\usepackage{hyperref}

\begin{document}

\title{Particle-in-cell simulations of  particle energization from low Mach number fast mode shocks} 
\author{Jaehong Park}
\affiliation{Department of Physics \& Astronomy, University of Rochester, Rochester NY, 14627}
\affiliation{Laboratory for Laser Energetics, University of Rochester, Rochester NY, 14623 }
\author{Jared C. Workman}
\affiliation{Department of Physics \& Astronomy, University of Rochester, Rochester NY, 14627}
\affiliation{Department of Physical \& Environmental Sciences, Colorado Mesa University, Grand Junction CO, 81501}
\author{Eric G. Blackman}
\affiliation{Department of Physics \& Astronomy, University of Rochester, Rochester NY, 14627}
\affiliation{Laboratory for Laser Energetics, University of Rochester, Rochester NY, 14623 }
\author{Chuang Ren}
\affiliation{Department of Physics \& Astronomy, University of Rochester, Rochester NY, 14627}
\affiliation{Laboratory for Laser Energetics, University of Rochester, Rochester NY, 14623 }
\affiliation{Department of Mechanical Engineering, University of Rochester, Rochester NY, 14627} 
\author{Robert Siller}
\affiliation{Department of Physics \& Astronomy, University of Rochester, Rochester NY, 14627}

\begin{abstract}
Astrophysical shocks are often studied in the high Mach number limit but 
weakly compressive  fast   shocks can  occur in magnetic reconnection outflows and are considered to be a site of particle energization in solar flares.  
Here we study  the microphysics of  such perpendicular, low Mach number collisionless shocks  using two-dimensional particle-in-cell (PIC) simulations with a reduced ion/electron mass ratio and employ a  moving wall boundary  method for initial generation of the shock. 
This moving wall method allows for more control of the shock speed, smaller simulation box sizes, and longer simulation times than the  commonly used fixed wall, reflection method of shock formation.
Our results, which are independent of the shock formation method, 
 reveal the prevalence  shock drift acceleration (SDA) 
 of both electron and ions in a purely perpendicular shock with  Alfv\'en Mach number $M_A=6.8$ and  ratio of thermal to magnetic pressure $\beta=8$.   
We determine the respective minimum energies required for  electrons and ions to incur SDA. We derive a theoretical electron distribution via SDA that compares to the simulation results. We also show that a modified two-stream instability due to the incoming and reflecting ions in the shock transition region  acts as the  mechanism to generate collisionless plasma turbulence that sustains the shock.
\end{abstract}

\maketitle

\section{Introduction}

Solar flares convert magnetic  energy  into flow and particle energy
via magnetic reconnection (see e.g., Priest and Forbes(2002)\cite{priest02}, Zharkova \textit{et al.}(2011)\cite{zharkova11}, and references therein). The outflows from such reconnection sites can exceed the fast magneto-sonic speed.  Unlike the inflows, the outflows from reconnection sites
are flow dominated and the ratio of thermal to magnetic pressure $(\equiv \beta)$ exceeds $1$. 
Analytic predictions\cite{blackman1} and numerical simulations in which an obstacle is present\cite{forbes88,workman11} reveal the presence of low Mach number fast shocks in these reconnection outflows.  In the standard geometry of a solar flare, such ``termination" shocks form, where the downward directed outflow interacts with the  magnetic loop formed from previously reconnected field lines.

Collisionless termination shocks have been invoked in a number of phenomenological solar flare models and may be associated with specific observational features. Mann \textit{et al.}(2006)(2009)\cite{mann06,mann09} suggested shock drift acceleration(SDA) as a mechanism of energetic electrons up to $\sim$ MeV during solar flares. Hard X-ray emission at loop top locations\cite{shibata95} may also be associated with such shocks. Decker and Vlahos(1986)\cite{decker86} carried out test particle simulations on the SDA in solar flare shocks.
Guo and Giacalone(2010)\cite{guo10} studied a solar flare shock using 
a hybrid simulation in the presence of turbulent magnetic fields, where electrons are efficiently accelerated by multiple reflection. The hybrid simulations treat the ions as particles and the electrons as a fluid.
However, to our  knowledge, fully-kinetic simulations, where both shock formation and particle acceleration are modeled from first principles in the regime of low-Mach-number and $\beta>1$ have not been reported.

Particle-in-Cell (PIC) simulations have been used to study high-Mach-number collisionless shocks and particle acceleration in ion-electron plasmas\cite{amano07,kato08,spitkovsky08,martins09,amano09,sironi11} but Low-Mach-number shocks are less widely studied.
Gargate and Spitkovsky(2012)\cite{gargate12} investigated a low Mach number shock as a subset of cases in a parameter survey using a hybrid code, where both diffusive shock acceleration (DSA) and SDA of the ion acceleration were observed, with the latter becoming more important as the shock becomes more perpendicular.

Here we report results from two-dimensional (2D) full-PIC simulations of perpendicular shocks in the regime of low-Mach-number  $(< 3)$ and high plasma $\beta (>1)$. The motivation is to study  the microphysics of shock formation and particle acceleration in the perpendicular shocks relevant to solar flares. Perpendicular shocks are chosen for their relevance to the shocks that emerge in the 2-D reconnection outflows\cite{workman11}
and for the first-step toward the study of more general quasi-perpendicular shocks. As we describe later, the simulations reveal both electron and ion acceleration via SDA. We also find that the modified two-stream instability by the interaction of incoming and reflecting ions in the shock transition region is the likely turbulent dissipation mechanism that
sustains the collisionless shock.

Most previous PIC simulations of collisionless shocks use a reflecting wall boundary condition where  plasma flow reflects off a rigid wall to form a shock. In that case, the simulation frame is fixed to the downstream rest frame. The shock moves away from the reflecting wall at a speed $V_{s}$, the downstream flow velocity in the shock rest frame. The simulation time is then  limited to $L_x/V_{s}$ where $L_x$ is the simulation box size in the direction of the shock propagation. 

In contrast our  simulations use a moving wall boundary condition, first introduced in Langdon \textit{et al.}\cite{langdon88}. It allows control of the downstream flow velocity and thus the shock speed in the simulation frame. By slowing down the shock speed in the simulation box, smaller boxes can be used for the same simulation time. We have implemented this boundary condition in 2D and find all properties of the generated shocks are similar to those with the reflecting boundary condition when differences between the reference frames are accounted for.

The rest of the paper is organized as follows. The simulation setup, including the moving wall boundary condition, is described in section \ref{setup}. The shock properties and particle acceleration are described in section \ref{result}. Discussion and summary are given in section \ref{summary}.

\section{Simulation Setup}
\label{setup}
\subsection{Basic setup}

We use the relativistic full PIC code OSIRIS\cite{osiris02} to study the formation of low Mach number fast perpendicular magnetosonic shocks and the particle acceleration therein. Following Refs.\cite{workman11} and \cite{tsuneta96}, typical parameters of solar flare reconnection outflows are  chosen as the upstream conditions for our shock. (Hereafter, ``upstream" is defined respect to the fast shocks we study herein, not upstream of a reconnection site.) 
In particular, we use a  plasma density $n=5\times 10^9$ cm$^{-3}$, electron and ion temperatures $T_e=T_i=0.5\text{keV}$, and the perpendicular magnetic field strength $B=5$G with $\beta\equiv 8\pi n(T_e+T_i)/B^2=8.05$. A reduced ion/electron mass ratio of $m_i/m_e=30$ is used to reduce the required computational resources. The Alfv\'en Mach number is chosen to be $M_A\equiv V_1\sqrt{4\pi m_in}/B=6.79$, which implies an upstream plasma flow velocity in the shock rest frame, $V_1=0.0274c$, where $c$ is the speed of light, for $m_i/m_e=30$. 
For a real proton-electron plasma,  
$V_1$ would be reduced by a factor of $\sqrt{30/1836}=0.128$ for the same $M_A$. The super-fast-magnetosonic Mach number $M$ satisfies  $M\equiv M_A/\sqrt{1+(5/6)\beta}=2.45$. The ratio of the electron cyclotron frequency to the electron plasma frequency is $\Omega_{ce}/\omega_{pe}=0.02207$.

With these upstream values of $M_A$ and $\beta$, the Rankine-Hugoniot relation\cite{tidman71} for perpendicular shocks gives the compression ratio $r=2.15$,
\begin{equation}
{V_{x2}\over V_{1}}={n_1\over n_2}={B_{z1}\over B_{z2}}\equiv{1\over r}=
{2\over 3M_{A1}^2}\left[1+\beta+{M_{A1}^2\over 2}\right]
\label{RHeq},
\end{equation}
where the lower indices $1$ and $2$ represent the upstream and the downstream, respectively, in the shock rest frame, and
we used the adiabatic index $\gamma=2$ for 2D.

A rectangular simulation domain in the $xy$ plane is used.  A uniform external magnetic field of $B=5$G is set out of the simulation plane (along the $z$-axis) and a uniform external $E$-field$(=V_dB/c)$ is set up along the $y$-axis. (An alternative in-plane configuration with $B_y$ and $E_z$ has also been used to help identify instabilities responsible for dissipation and has yielded similar results.) The simulation box is initialized with a Maxwellian ion-electron plasma drifting to the right with $V_d=0.021c$ and $T_e=T_i=0.5\text{keV}$,
where $V_d$ is set to a smaller value than the upstream speed $V_1(=0.0274c)$ in the shock rest frame considering the shock speed traveling to the left direction.
A new plasma of the same distribution is constantly injected from the left boundary ($x=0$) throughout the simulation.  The simulation box sizes are $L_x\times L_y=340c/\omega_{pe} \times 40 c/\omega_{pe}$. These correspond to $62c/\omega_{pi}\times 7.3 c/\omega_{pi}$, where $\omega_{pi}$ is the ion plasma frequency, for $m_i/m_e=30$. The grid size used is $dx=dy=0.08c/\omega_{pe}$ and the time step used is $dt=0.056/\omega_{pe}$. For diagnostic purposes,  a small population of electrons and ions spatially localized within a circular region, is set as a separate species for which the particle information is stored more frequently to track  detailed trajectories over time. For each particle species, 25 particles per cell are used. A linear current deposition scheme is used for all simulations in this paper.

A periodic boundary condition is used in the $y$-direction for both particles and fields. For fields, an open boundary condition is used in the $x$-direction. Particles that reach  $x=0$ are re-injected into the box with the initial drifting Maxwellian distribution. At $x=L_x$, a moving wall boundary condition is adopted, as described in the next subsection.

\subsection{The moving wall boundary condition}
In general, a reflecting wall moving in the direction of the flow\cite{langdon88} can force the plasma flow velocity at the wall to be an arbitrary predetermined velocity by selectively reflecting particles with certain velocities. Here we implement a moving wall boundary condition at $x=L_x$ to force the flow velocity there to be close to the downstream velocity measured in the shock frame. The wall is essentially treated as an infinitely massive slab moving with  velocity $v_{\text{wall}}$ and particles that catch up to it rebound specularly off the wall in the wall rest frame. Proper relativistic momentum transformations are applied to obtain the particle velocity in the simulation frame after rebounding.
Between  times $t_{n}$ and $t_n+\Delta t$, the wall will have ``moved" a distance $V_{\text{wall}}\Delta t$ from the right domain boundary $L_x$.  Particles will also be moving beyond the domain boundary during this time.  Particles that can reach the wall and rebound quickly enough to return to the simulation domain during $\Delta t$ are kept in the box.  Those which do not are removed from the system. At the right boundary, this procedure forces the bulk flow velocity to be $V_{\text{wall}}$. 

One immediate consequence of this implementation is that it cannot be used to probe shocks with compression ratios of $r=2$ or less. This can be seen by noting that a necessary condition for the plasma to return to the simulation domain is that its updated velocity must be negative in sign.  In one dimension, a  non-relativistic particle with a velocity $v_p$ will rebound with a velocity $2V_{\text{wall}}-v_p$. This must be negative for the particle to remain in the box.
Initially, $v_p\approx V_d$ where $V_d$ is the plasma flow velocity from the left boundary. In the shock rest frame, $V_d$ is the upstream velocity and $V_{\text{wall}}$ should be the downstream velocity, $V_{\text{wall}}=V_d/r$.  The condition for rebound back into the box then becomes $2V_d/r -V_d < 0$, where $r=n_2/n_1$ is the compression ratio across the shock. 
For $r \le 2$ the rebound condition cannot be satisfied for purely elastic collisions with the wall.

In practice we do not set the wall velocity to be exactly the downstream velocity in the shock frame but set it so that the shock is slowly propagating back into the upstream.  This is necessary to allow a large enough downstream region to be generated wherein particles may travel several ion gyro-radii to undergo acceleration. We find that we can control the shock velocity in the simulation frame by changing $V_{\text{wall}}$ and have tested that the properties of these shocks are essentially the same as those  generated from a stationary reflecting wall boundary.

\section{Results and analysis}
\label{result}
\subsection{shock structure}

\begin{figure}
\includegraphics[scale=.300]{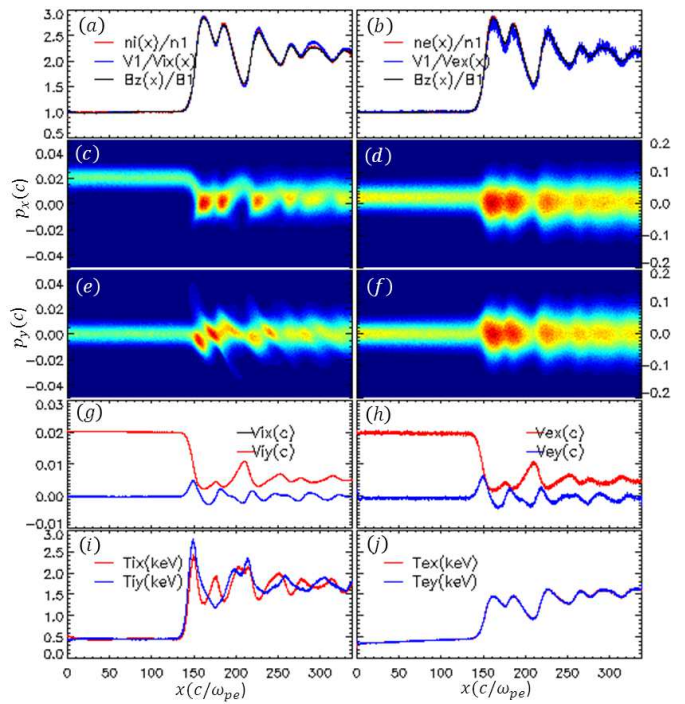}
\caption{(Color online) The ratios, $n/n_1$, $V_1/V_x$, and $B_z/B_1$,
momentum distribution of $p_x$ and $p_y$, $y$-averaged flow velocity of $V_x$ and $V_y$, and temperature $T_x$ and $T_y$, for ions(left column) and electrons(right column) at $t=28476/\omega_{pe}$.
$(a)\sim(b)$ are calculated in the shock-rest frame and the other plots are obtained in the simulation frame.}
\label{phase}
\end{figure}

In Fig.\ref{phase}, we plot the ratios of the density, velocity, and magnetic field [$n(x)/n_1$, $V_1/V_x(x)$, and $B_z(x)/B_1$], the momentum phase space distributions of $p_xx$ and $p_yx$, the $y$-averaged flow velocity profiles $V_x(x)$ and $V_y(x)$, and the temperature profiles $T_x(x)$ and $T_y(x)$, for the ions (the left column) and the electrons (the right column) at $t=28476/\omega_{pe}$. Here, the momenta are defined as $p_x=\Gamma v_x$ and $p_y=\Gamma v_y$, where $\Gamma=1/\sqrt{1-(v/c)^2}$. The ratio curves [(a) and (b)] are calculated in the shock-rest frame using the Lorentz transformation and the other plots are computed in the simulation frame.

In Fig.\ref{phase}, the shock front is at $x\approx 150c/\omega_{pe}$ and moves to the left with a shock speed in the simulation frame of $6.4\times 10^{-3}c$. An oscillatory pattern in the downstream properties can be observed, which indicates weak dissipation in this low-Mach-number shock. The Rankine-Hugoniot condition in Eq.(\ref{RHeq}) gives $r=2.15$ for our upstream parameters. In the simulation, the compression ratio is $r\sim 2.8$ near the shock front and relaxes to $r\sim 2.1$ in the far downstream with a weakly damped oscillatory pattern (Fig.\ref{phase}a and b). In the downstream, the electrons are thermalized isotropically (Fig.\ref{phase}d, f and j) but the ions are heated slightly more in the $y$-direction near the shock front (Fig.\ref{phase}e and i). The electrons are heated to $T_2\approx 1.4\text{keV}$ and $T_2/T_1\approx 2.8$ (Fig.\ref{phase}j), which indicates that the electrons are mainly heated by adiabatic compression, $(T_2/T_1)_{adia}=(n_2/n_1)^{\gamma-1}=r^{\gamma-1}=r$, where the adiabatic index $\gamma$ is 2 in 2D. The electron temperature in the downstream slightly increases as the compression ratio relaxes to $\sim 2.1$ (Fig.\ref{phase}j).
The ions are heated to a value much higher than the electrons near the shock front (namely $T_2/T_1\approx 3\sim 5$ in Fig.\ref{phase}i), highlighting the more substantial role of non-adiabatic particle energization for the ions than the electrons.

The flow velocity in the downstream is mainly in the $x-$direction, oscillating around $V_\text{{wall}}(=0.0052c)$ (Fig.\ref{phase}g and h). The flow velocity in the $y$-direction oscillates around zero in the downstream (Fig.\ref{phase}g and h) due to the ${\bf E}_x\times {\bf B}$-drift. The damped oscillatory pattern in the downstream results from weak turbulent dissipation in the shock transition region via the modified two-stream instability\cite{papadopoulos71,wagner71}. This instability arises from the incoming and reflecting ions at the shock front. The ions are reflected in a thin region, $140<x<155(c/\omega_{pe})$ (Fig.\ref{phase}c), due to a potential jump, $e\Delta\Phi$, across the shock front given by the electron momentum equation\cite{leroy82,hoshino01},
\begin{align}
e\Delta\Phi
&\approx\int_{-\infty}^{x}{1\over n}{\partial nT_e\over\partial x}dx
+\int_{-\infty}^{x}{B_1\over 4\pi n_1}{\partial B_z\over\partial x}dx 
\nonumber \\
&= 2T_1(r-1)+{(r-1)\over M_A^2} m_iV_1^2= 2.3\text{keV},
\label{potential_eq}
\end{align}
where the ion drift $V_{iy}$ is omitted.

\begin{figure}
\includegraphics[scale=.31]{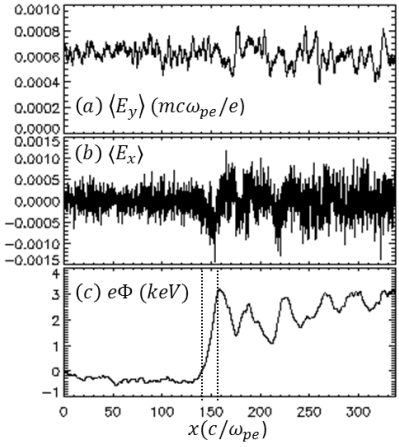}
\caption{$y$-averaged $E_y$, $E_x$, and the potential energy $e\phi(x)$ at $t=28476/\omega_{pe}$ in the shock rest frame. In (c), the shock transition region is indicated by the dotted lines.}
\label{field}
\end{figure}

Figure \ref{field} shows the $y$-averaged $E_y$, $E_x$, and the electric potential energy $e\Phi(x)$ at $t=28476/\omega_{pe}$ in the shock rest frame. $E_y$ is positive and approximately constant across the shock (Fig.\ref{field}a), but  a negative $E_x$ causes a potential barrier for the incoming ions at the shock front (Fig.\ref{field}b and c). The potential energy is $e\Phi(x)\approx 3$keV
[a bit larger than the $2.3$keV from Eq.(\ref{potential_eq})]
and can reflect the low energy tail of the incoming ions, which have an average drift energy in the shock rest frame of $5.74$keV ($V_1=0.0274c$).

\subsubsection{Modified Two-Stream Instability as the Source of Dissipation for Shock Sustenance}
\begin{figure}
\includegraphics[scale=.25]{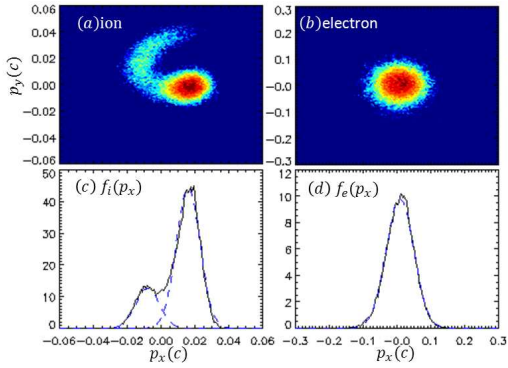}
\caption{(Color online) (a)(c)ion and (d)(d)electron distributions in the shock transition region, $140<x<155(c/\omega_{pe})$ in the simulation frame.
In (c) and (d), the distributions fit into Maxwellian distributions(dashed lines).}
\label{fdist}
\end{figure}

\begin{figure}
\includegraphics[scale=.32]{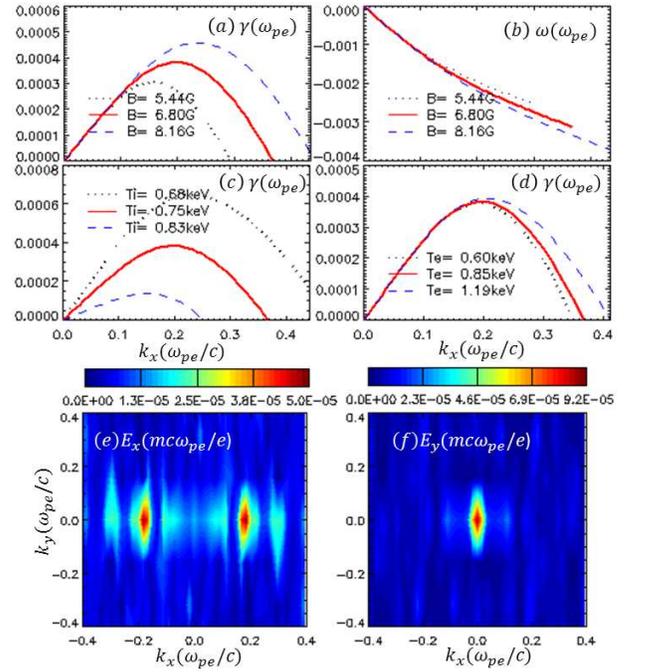}
\caption{(Color online) 
(a)$\sim$(d): Numerical solutions of the modified two-stream instability under $B=6.8$G, $T_i=0.75$keV, and $T_e=0.85$keV (solid lines), and different parameters (dotted and dashed lines). 
(e) and (f): Fourier spectra of $E_x$ and $E_y$ fields from the simulation
at $t=28476/\omega_{pe}$.}
\label{mt30}
\end{figure}

Figure \ref{fdist} shows the electron and ion distributions in the shock transition region, $140<x<155(c/\omega_{pe})$. The ions have a bump-on-tail distribution, with $22\%$ of the ions reflected (Fig.\ref{fdist}a and c). A high energy component can also be observed moving in the positive-$y$ direction (Fig.\ref{fdist}a), resulting from SDA, which will be discussed later. The electrons are essentially isotropic and drift with $V_{ex}=0.0085c$(Fig.\ref{fdist}b and d). We use the distribution in Fig.\ref{fdist}(c) and (d) for a  linear stability analysis of the modified two-stream instability(MTSI) at the shock transition region in the  electron drift rest frame to assess whether the associated dissipation is consistent with 
what is needed to sustain the shock.

For the stability analysis, we also use the following initial parameters extracted directly from the simulation: The magnetic field is $B_z=1.36B_1=6.8$G. The electrons are magnetized and have a Maxwellian distribution with a temperature of $T_\text{e}=0.85$keV. The ions are assumed to be non-magnetized and have drifting Maxwellian distributions. In the electron rest frame, the drift velocities and densities for the incoming and reflecting ions are $V_{x\text{in}}=0.0075c$, $V_{x\text{re}}=-0.0165c$, $n_\text{in}=0.78n_i$, $n_\text{re}=0.22n_i$. Both incoming and reflecting ions have a temperature of $T_{x\text{in}}= T_{x\text{re}}=0.75$keV.

\begin{figure}
\includegraphics[scale=.28]{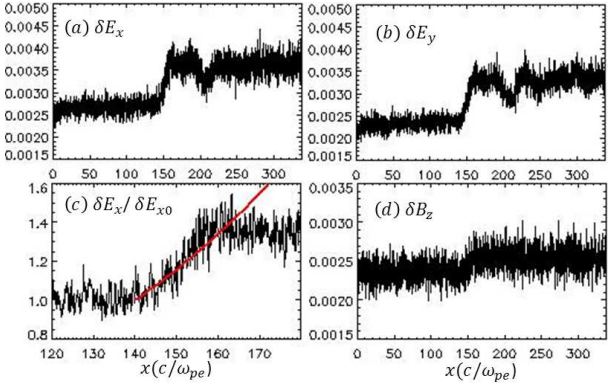}
\caption{(Color online) field fluctuation, $\delta E_x$, $\delta E_y$, and $\delta B_z$ at $t=28476/\omega_{pe}$. In (c), $\delta E_x/\delta E_{x0}$ is plotted and
the red line is $\text{exp}[\gamma (x-x_0)/V_1]$, where $\gamma=4\times 10^{-4}\omega_{pe}$ from the modified two-stream instability, $x_0=140c/\omega_{pe}$,
and $V_1=0.0274c$.}
\label{fluctuation}
\end{figure}

The dispersion relation for the MTSI with $\vec{k}=k\hat{x}$ is\cite{gary93}
\begin{eqnarray}
&&1+{\omega_{pe}^2\over k^2 v_{eth}^2}
\left(1-e^{-\lambda_e}\sum_{m=-\infty}^{\infty}I_m(\lambda_e)
{\omega\over\omega-m\Omega_{ce}}\right) \nonumber \\
&&-\sum_{s=\text{in},\text{re}}{\omega_{ps}^2\over 2 k^2 v_{sth}^2}Z'(\xi_s)=0,
\label{mtdisp}
\end{eqnarray}
where $v_{sth}(=\sqrt{T_s/m_s})$ is the thermal velocity of species $s(=e,\text{in},\text{re})$, $\lambda_e=k^2 v_{eth}^2/\Omega_{ce}^2$, $I_m(\lambda_e)$ is a modified Bessel function of the 2nd kind, $\xi_s=(\omega-kV_{xs})/\sqrt{2}kv_{sth}$, and $Z(\xi)$ is the plasma dispersion function. We then numerically solve Eq.(\ref{mtdisp}), using a fractional polynomial approximation of the $Z(\xi)$ function\cite{nakamura98}, the Zenkins and Traub algorithm\cite{jenkins70,jenkins72} for complex polynomial root finding, and the Muller method\cite{bose09} to obtain accurate numerical solutions.

Figure \ref{mt30} shows numerical solutions of Eq.(\ref{mtdisp}) for the parameters given in the paragraphs above (solid lines) and also for slightly different parameters (dotted and dashed lines).
The maximum growth rate in the solid line is $\gamma=4\times 10^{-4} \omega_{pe}$ at $k=0.2\omega_{pe}/c$ (Fig.\ref{mt30}a). 
The growth rate increases as the magnetic field increases (Fig.\ref{mt30}a) and decreases as the ion and/or electron temperatures increase (Fig.\ref{mt30}c and d), but is less sensitive to $T_e$ than to $T_i$. In Fig.\ref{mt30}(b), the real frequency is negative, giving a phase velocity comparable to the drift velocity of the reflected ions in the electron rest frame. We compare the numerical solutions with the modes observed in the simulation: In Fig.\ref{mt30}(e) and (f), the Fourier spectra of $E_x$ and $E_y$ fields from the simulation are plotted. An electrostatic mode is observed at $k_x=0.2 \omega_{pe}/c$ in $E_x$, in agreement with the MTSI dispersion relation.

Figure \ref{fluctuation} shows the fluctuating fields along the $x$-axis
at $t=28476/\omega_{pe}$ for $\delta E_x$, $\delta E_y$, and $\delta B_z$, where we define 
$\delta A\equiv\sqrt{\langle (A-\langle A\rangle)^2\rangle}$ and $\langle..\rangle$ is the ensemble average. Here we use the $y$-average
as the ensemble average. 
In (a) and (b), $\delta E_x$ and $\delta E_y$ increase at the same rate in the shock transition region.
In (c), we plot the ratio of $\delta E_x$ to $\delta E_{x0}$, where $\delta E_{x0}$ is the fluctuating field in the upstream. 
In (d), the magnetic field fluctuation is negligible compared to the electric field fluctuation since the MTSI has electrostatic modes.
The red line in (c) is the evolution curve of the fastest growing mode in the MTSI, $\text{exp}[\gamma (x-x_0)/V_1]$, 
where the growth rate $\gamma=4.\times 10^{-4}\omega_{pe}$ and the shock speed $V_1=0.0274c$.
The MTSI begins to operate at $x=x_0=140c/\omega_{pe}$ and ends before 
$x=160c/\omega_{pe}$ during the shock transit time, $\Delta t=\Delta x/V_1\sim 730/\omega_{pe}$, and is enough to excite the 
electric field fluctuation observed in the shock transition region.

Whether this level of $\delta\bold{E}$ is sufficient to generate the entropy increase required for the shock is an interesting question. The entropy equation can be written as by acting $\int d\bold{v}(1+\text{ln}\langle f\rangle)$ on the Vlasov equation \cite{tidman71}\cite{parks12}
\begin{eqnarray}
\partial(ns)/\partial t+
\partial(nV_xs)/\partial x=
\partial/\partial x\int d\bold{v}'v_x'\langle f\rangle\text{ln}\langle f\rangle \nonumber \\
+{e\over m}\int {d\bold{v}\over\langle f\rangle}
\left\langle{\partial\langle f\rangle\over\partial\bold{v}}\cdot
\left(\delta\bold{E}+{1\over c}\bold{v}\times\delta\bold{B}\right)\delta f \right\rangle
\label{entropy_eq1},
\end{eqnarray}
where $s(x)$ is the specific entropy (entropy/particle),  
$s(x)=-\int d\bold{v}\{\langle f(x,\bold{v})\rangle\
\text{ln}\langle f(x,\bold{v})\rangle\}/\int d\bold{v}\langle f(x,\bold{v})\rangle$,
$n=\int d\bold{v}\langle f\rangle$, $V_x=(1/n)\int d\bold{v}v_x\langle f\rangle$, and $v_x'=v_x-V_x$. In the shock rest frame $(\partial/\partial t=0)$, Eq.(\ref{entropy_eq1}) can be written as
\begin{eqnarray}
s(x)-s_1-{F(x)-F_1\over n_1V_{x1}} 
={1\over n_1V_{x1}}\int_{1}^{x} dx \nonumber \\
{e\over m}\int {d\bold{v}\over\langle f\rangle}
\left\langle{\partial\langle f\rangle\over\partial\bold{v}}\cdot
\left(\delta\bold{E}+{1\over c}\bold{v}\times\delta\bold{B}\right)\delta f
\right\rangle.
\label{entropy_eq2}
\end{eqnarray}
Here $F(x)=\int d\bold{v}'v_x'\langle f\rangle\text{ln}\langle f\rangle$
as seen in Ref.\cite{parks12}.

\begin{figure}
\includegraphics[scale=.33]{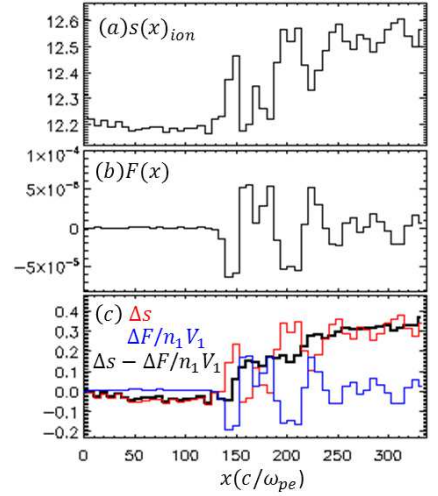}
\caption{(Color online) (a) specific entropy $s(x)$ for ions
(b) $F(x)$ for ions (c) $\Delta s=s(x)-s_1$ (red), 
$\Delta F/n_1V_{x1}=(F(x)-F_1)/n_1V_{x1}$ (blue),
$\Delta s-\Delta F/n_1V_{x1}$ (black) for ions.}
\label{entropy}
\end{figure}

In principle, one could measure every term in Eq.(\ref{entropy_eq2}) and show definitively whether this level of $\delta\bold{E}$ is sufficient to generate the entropy increase required. However, the term on the right-hand side, the entropy change due to fluctuating fields, involves $\delta f$ and is difficult to evaluate from the simulation. Here, we measure the terms on the left-hand side of Eq.(\ref{entropy_eq2}).
Figure \ref{entropy}(a) shows the specific entropy for the ions measured from the raw particle data in the simulation. 
The specific entropy jump between the upstream and downstream is $\sim 0.4$.
The Sackur-–Tetrode equation for the entropy of an ideal gas gives a similar result for the entropy jump in this simulation, 
$\Delta s=(S_2-S_1)/N=\text{ln}\left\{(1/r)(T_2/T_1)^{d/2}\right\}=0.35$,
where $d$ is the degree of freedom ($d=2$ for 2D), $r=2.1$, and $T_2/T_1=1.5/0.5$.
Figure \ref{entropy}(b) shows $F(x)$  for the ions in Eq.(\ref{entropy_eq2}).  When $\langle f(\vec{v})\rangle$ is an even function about $V_x$ such as a drift-Maxwellian distribution, $F(x)$ vanishes. We see non-zero $F(x)$ in the transition region. 
In Fig.\ref{entropy}(c), we plot $\Delta s=s(x)-s_1$ (red line), 
$\Delta F/n_1V_{x1}=\{F(x)-F_1\}/n_1V_{x1}$ (blue line), 
and $\Delta s-\Delta F/n_1V_{x1}$ (black line) for the ion.
Therefore, we conclude that the fluctuating fields excited by the MTSI are
necessary for the entropy creation throughout the downstream region. Whether they are sufficient remains an open question.


Another possible instability responsible for shock formation comes from diamagnetic currents on the shock front\cite{tidman71}, and would have a mode with $k_y$. This mode, however, is not observed in either $E_x$- or $E_y$-spectrum as seen in Fig.\ref{mt30}(e,f).
This is supported by another perpendicular shock simulation where the magnetic field is initiated in the simulation plane and the $k_y$-mode is precluded but the same shock structure is seen.

\subsection{Particle heating via shock drift acceleration}
\subsubsection{Dynamics of ions and electrons incurring SDA}

\begin{figure}
\includegraphics[scale=0.33]{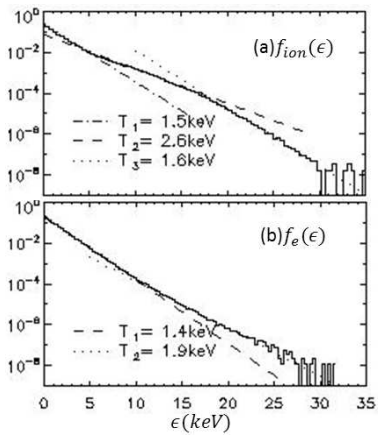}
\caption{
Simulation result of the energy distribution,
$f(\epsilon)=(1/N_\text{tot})dN(\epsilon)/d\epsilon$,
in the downstream rest frame of $145<x<340(c/\omega_{pe})$ at $t=28476/\omega_{pe}$.
(a) Ion energy distribution and fittings into Maxwellian temperatures
with $T_1=1.5$(dot-dashed), $2.6$(dotted), and $1.6$(keV)(dotted lines).
(b) Electron energy distribution and fitting into Maxwellian temperatures 
with $T_1=1.4$keV(dashed line) in $0<\epsilon<10$ and $T=1.9$keV(dotted line)
in $\epsilon>13$(keV).}
\label{edist}
\end{figure}

Shock surfing acceleration
(SSA)\cite{hoshino01,hoshino02,amano07,kato08,dieckmann06}, whereby particles are trapped in the solitary wave structure excited by the Bunemann type instability at the shock front and then accelerated by the convective $E_y$ field along the shock, is inefficient for low Mach perpendicular shocks. We find no evidence of SSA in our simulation. 
However, when  particles pass through the shock transition region, the magnetic field jump at the shock front allows particles to experience shock drift acceleration (SDA)\cite{webb83,decker86,whipple86,decker88,begelman90,kirk94,ball01}. SDA results from the fact that the $E_y$ is constant across the shock but magnetic field jump gives different gyro-radii of a particle at two sides of the shock. 
This gives rise to a net drift along the $y$-axis, $\delta y$, and a net gain energy of $\delta\epsilon=eE_y\delta y$ per particle.

In our simulation, both ion and electron heating via SDA are observed. Figure \ref{edist} shows the  normalized ion and electron distributions in the downstream rest frame,
$f(\epsilon)=(1/N_\text{tot})dN(\epsilon)/d\epsilon$,
in the downstream region of $145<x<340(c/\omega_{pe})$ at $t=28476/\omega_{pe}$. Both distributions have a low energy regime that corresponds to adiabatic heating($\epsilon\lesssim 5$keV for the ion and $\epsilon\lesssim 10$keV for the electron) and a high energy regime due to SDA. 
The ion distribution fits into a multi-temperature Maxwellian distribution with temperatures of $T_{i1}=1.5$keV in $0<\epsilon<5$(keV), $T_{i2}=2.6$keV in $5<\epsilon<17$(keV), and $T_{i3}=1.6$keV in $\epsilon>17$keV (Fig.\ref{edist}a). 
The electron distribution fits into a two-temperature Maxwellian distribution with temperatures of $T_{e1}=1.4$keV in $0<\epsilon<10$(keV),
$T_{e2}=1.9$keV in $\epsilon>13$(keV) (Fig.\ref{edist}b).
The minimum energy where SDA is effective has different origins for the ions and electrons. The difference mainly comes from the fact that the magnetic moment is conserved for the electrons but not for the ions in the shock transition region.

\begin{figure}
\includegraphics[scale=.29]{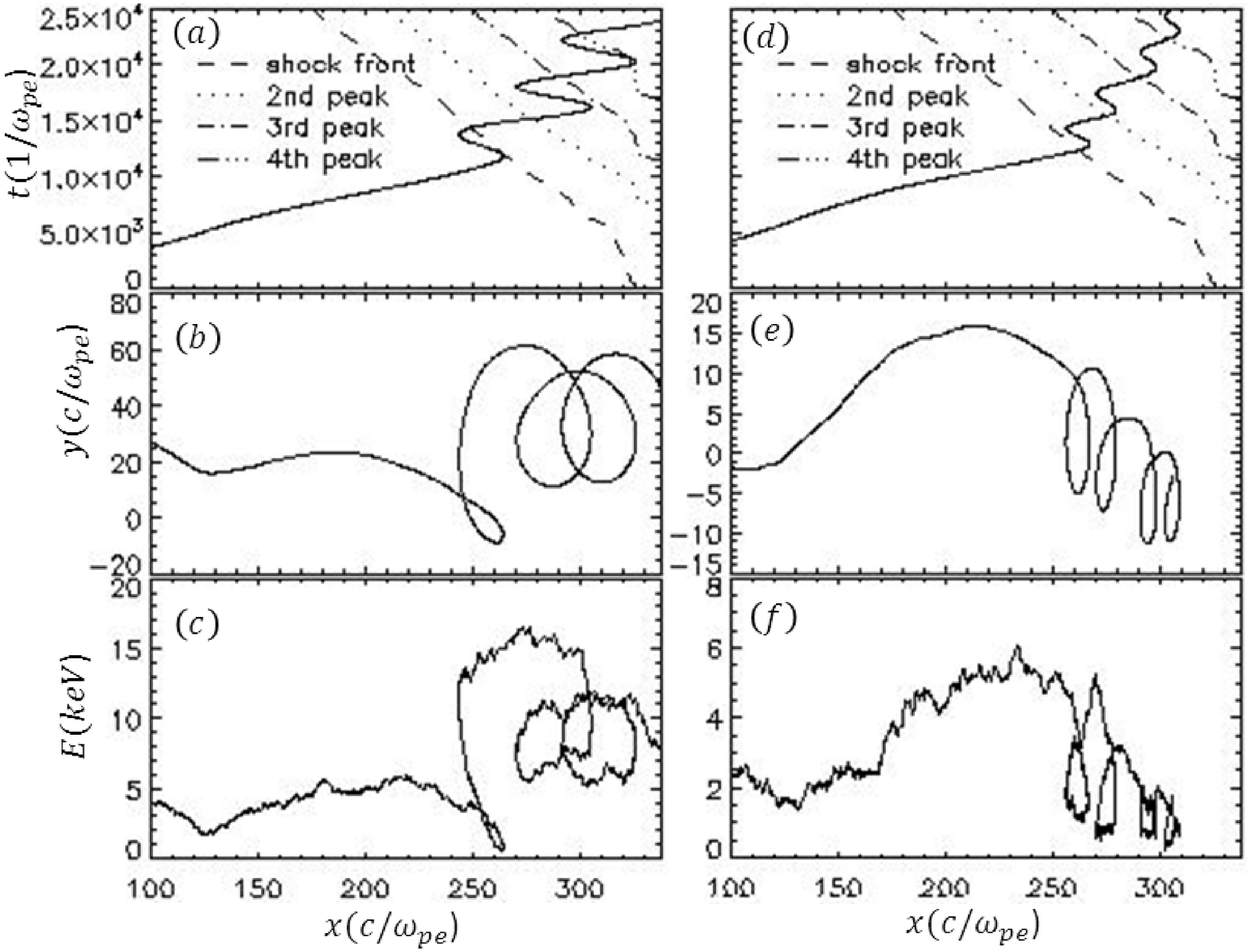}
\caption{A typical ion tracking experiencing SDA (in the left column) and not experiencing SDA (in the right column). In (a), we plot the positions of the shock front, the 2nd, 3rd, and 4th peaks traveling to the $-x$ direction in the simulation frame.}
\label{itracking}
\end{figure}

We discuss SDA for the ions first. Figure \ref{itracking} shows typical tracks for an ion experiencing SDA (the left column) and for one not experiencing SDA (the right column) from the simulation. Figure \ref{itracking}(a) shows the particle's $x$-coordinate vs time, along with positions of the shock front and the subsequent 3 compression peaks identifiable in Fig.\ref{phase}(a). When the particle meets the shock front, it turns back toward the upstream. The trajectory in the $xy$-plane in Fig.\ref{itracking}(b) shows the particle drifting up along the $y$-axis with a larger gyro-radius after turning back. Accelerated by the $E_y$ field, the kinetic energy of the ion increases from $4$keV up to $16$keV after encountering the shock front [Fig.\ref{itracking}c]. This particle later re-crosses the shock front and passes through the secondary compression peaks where it loses  some energy, reaching the right boundary with an energy of $8$keV. The net energy gain of the particle results from SDA at the encounter with the shock front, despite some energy loss during interaction with the secondary peaks.
In Fig.\ref{itracking}(d)-(f), a different ion entering the shock front with a different angle and energy does not meet the required conditions (discussed below) to turn back to cross the shock front again and the kinetic energy of the ion decreases to $1$keV after reaching the right boundary. 

\begin{figure}
\includegraphics[scale=.34]{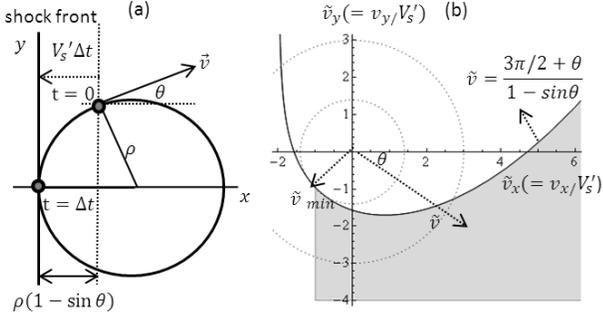}
\caption{
(a) A schematic view of the ion trajectory in the flow rest frame in the shock transition region where the shock front travels to the left with speed $V_s'$.
(b) The shaded region represents the ions experiencing SDA in the flow rest frame. The minimum velocity to experience SDA is indicated by $\tilde{v}_\text{min}(=v/V_s')$.}
\label{sda_sol}
\end{figure}

Whether or not an ion gains energy via SDA in a perpendicular shock depends on its incident speed and angle of incidence at the shock front,
and this is similar to the case of the electrons in a relativistic perpendicular shock\cite{begelman90,kirk94}.
The minimum energy above which SDA is operative is best computed in a reference frame where the ion drift velocity nearly vanishes and the ion executes approximately pure gyro-motion. Because the upstream and downstream drift velocities are different, the reference frame is chosen such that the shock front moves to the left with speed $V'_s=1/2(V_1+V_2)=V_1(1+r)/(2r)$ (Fig.\ref{sda_sol}a). Since the gyro-radius for an ion is larger in the upstream than in the downstream, an ion will only gain energy if it returns to the upstream side drifting upward. A necessary condition for the ion to return to the upstream side, as shown in Fig.\ref{sda_sol}(a), is
\begin{equation}
\rho(1-\text{sin}\theta)> V'_s\Delta t,\,\,\,
v\text{cos}\theta>-V'_s,
\label{sda_cond1}
\end{equation}
where $\rho= v/\Omega_{c}$ is the gyro-radius ($v>0$ is the particle speed and $\Omega_{c}$ is the gyro-frequency), and $\Delta t=(3\pi/2+\theta)/\Omega_{c}$ is the time between the shock front crossings.
Equation (\ref{sda_cond1}) can be re-written as
\begin{equation}
\tilde{v}>{3\pi/2+\theta\over 1-\text{sin}\theta},\,\,\,
\tilde{v}>-{1\over\text{cos}\theta},
\label{sda_cond2}
\end{equation}
with a dimensionless variable $\tilde{v}=v/V_s'$.
In Fig.\ref{sda_sol}(b), Eq.(\ref{sda_cond2}) is plotted in the $\tilde{v}_x\tilde{v}_y$-plane 
where the ions in the shaded region satisfy Eq.(\ref{sda_cond2}). 
The minimum velocity for the ions that can gain energy through SDA 
is indicated by the shortest distance from the origin to the shaded area in Fig.\ref{sda_sol}(b) and is measured to be $\tilde{v}_\text{min}=1.38$. The corresponding minimum energy is $\epsilon_\text{min}=1/2M(\tilde{v}_\text{min}V'_s)^2$=5.04 keV, using the measured value of $r=2.8$. This agrees reasonably well with the transition at $\epsilon_\text{min}\sim 5$keV in Fig.\ref{edist}(a), measured in the downstream rest frame. 

\begin{figure}
\includegraphics[scale=.29]{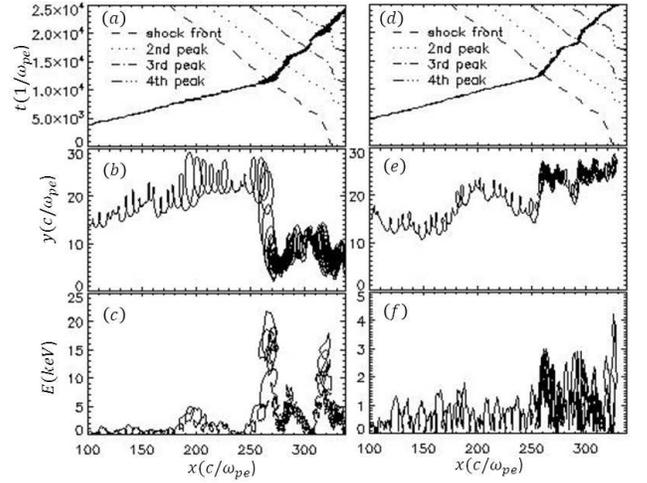}
\caption{A typical electron tracking experiencing SDA (in the left column) and not experiencing SDA (in the right column) in the simulation frame.}
\label{etracking}
\end{figure}

We now describe electron SDA. In Fig.\ref{etracking}, we plot typical tracks of an electron experiencing SDA (the left column) and one not experiencing SDA 
(the right column) from the simulation. Unlike the ions in Fig.\ref{itracking}, the electrons drift through the shock (Fig.\ref{etracking}a and d) without turning back. Since the electron gyro radius is small compared to the shock width, the electron magnetic moment can be treated as a constant in the shock transition region. In addition to the drift in the $x$-direction, electrons also drift in the $y$-direction due to ${\bf E}_x\times {\bf B}$-drift and $\nabla {\bf B}$-drift.

In Fig.\ref{etracking}(b), the electron drifts downward $(-\hat{y}$ axis) due to the $\nabla {\bf B}$-drift in the shock transition region around $x= 260c/\omega_{pe}$ and its kinetic energy increases up to $20$keV after encountering the shock front [Fig.\ref{etracking}b]. After leaving the shock front, the energy decreases to $1$keV in the $\partial B/\partial x<0$ region and then increases again to $17$keV once it encounters the 4th compression peak.
In contrast, the electron in Fig.\ref{etracking}(e) drifts upward 
$(\hat{y}$ axis) due to the ${\bf E}_x\times {\bf B}$-drift($E_x<0$ in the shock transition region) and its energy is fluctuating between $0<\epsilon<3$ keV.

For an electron to experience SDA, it must drift downward, which is possible when the $\nabla{\bf B} $-drift is larger than the ${\bf E}_x\times {\bf B}$-drift in the shock transition region where $\partial B/\partial x>0$ and $E_x<0$. This condition can be written as
\begin{equation}
{mv^2c\over 2eB_z^2}{\Delta B_z\over\Delta x}>{c|E_x|\over B_z},
\end{equation}
where $\Delta x$ is the width of the shock transition region.
The minimum energy for the electrons to experience SDA is then
\begin{equation}
\epsilon_\text{min}=e|E_x|B_z{\Delta x\over\Delta B_z}=
{e|E_x|\Delta x\over 2}\left({r+1\over r-1}\right),
\label{emin}
\end{equation}
where $B_z$ and $\Delta B_z$ are taken as 
$(r+1)B_1/2$ and $(r-1)B_1$, respectively.
With $r=2.8$ and $E_x=-0.0003$ $mc\omega_{pe}/e$ for our shock, $\epsilon_\text{min}$ is estimated to be $3.24$keV.
The negative $E_x$ field in the shock transition region 
requires the electrons to have higher threshold energy for SDA.

\subsubsection{Electron Spectrum from SDA}

Now we discuss the electron spectrum due to SDA. 
Neglecting adiabatic heating, the electron energy change during one gyro-cycle  in the shock transition region is $\delta\epsilon_\text{SDA}=eE_y\delta y$, where $\delta y$ is the net drift distance from the $\nabla {\bf B}$-drift and the ${\bf E}_x\times {\bf B}$-drift during one gyro-cycle and is given by
\begin{equation}
\delta y=\left({\epsilon c\over eB_z^2}{\Delta B_z\over\Delta x}-{c|E_x|\over B_z}\right){2\pi\over\Omega_c}.
\end{equation}
Here, $\epsilon(=mv^2/2)$ is the electron energy of gyro-motion.
The rate of energy change from SDA is given by
\begin{equation}
{d \epsilon_\text{}\over d t}={\alpha\epsilon-\eta\over\tau},
\label{esda}
\end{equation}
where $\tau=2\pi/\Omega_c$ is the period of one gyro-cycle, and 
$\alpha$ and $\eta$ are defined respectively as
\begin{equation}
\alpha\equiv{2\pi mc V_1 B_1\over eB_z^3}{\Delta B_z\over\Delta x},\,
\eta\equiv{2\pi mc V_1 B_1 |E_x|\over B_z^2}.\,
\label{constant}
\end{equation} 
The solution to Eq.(\ref{esda}) is
\begin{equation}
\epsilon=(\epsilon_1-\epsilon_\text{min})
\text{exp}\left[{\alpha t\over \tau}\right]+\epsilon_\text{min},
\,\,\,(\epsilon_1>\epsilon_\text{min}),
\label{esda_sol}
\end{equation}
where $\epsilon_1$ is the particle energy before entering the shock transition region and $\epsilon_\text{min}$ is written as 
$\epsilon_\text{min}=\eta/\alpha$ using Eq.(\ref{emin}) and (\ref{constant}).

If we assume that the particle enters the shock at a time $t=0$ with energy $\epsilon_1$
then the probability for a single electron to leave the shock transition region 
at a time $t$ is given by a Dirac delta function,
\begin{equation}
P(t)dt=\delta(t-\Upsilon)dt,
\label{delta}
\end{equation}
where $\Upsilon$ is a characteristic time for escape and is
given by $\Upsilon=\Delta x/\langle V_x\rangle$, 
$\Delta x$ is the width of the shock transition region, and $\langle V_x\rangle$ is the averaged drift velocity. The associated probability of an electron to change its energy from $\epsilon_1$ to $\epsilon$ is
\begin{equation}
P(\epsilon,\epsilon_1)d\epsilon=
\delta\left[\epsilon-(\epsilon_1-\epsilon_\text{min})e^{\alpha\Upsilon/\tau}
-\epsilon_\text{min}\right]d\epsilon.
\end{equation}
The energy distribution in the downstream rest frame is obtained by
\begin{equation}
f_2(\epsilon_2)=\int_{\epsilon_\text{min}}^{\infty}d\epsilon_1 f_1(\epsilon_1)P(\epsilon_2,\epsilon_1),
\label{fdown}
\end{equation}
where $f_1(\epsilon_1)$ is the energy distribution in the upstream rest frame and given by $(1/T_1)e^{-\epsilon_1/T_1}$. 
Then the energy distribution after SDA is 
a translated Maxwellian distribution with a temperature $T_2=T_1e^{\alpha\Upsilon/\tau}$ for $\epsilon_2>\epsilon_\text{min}$,
\begin{equation}
f_2(\epsilon_2)={1\over T_1 e^{\alpha\Upsilon/\tau}}
\text{exp}\left[-{1\over T_1e^{\alpha\Upsilon/\tau}}
\left\{\epsilon_2+(e^{\alpha\Upsilon/\tau}-1)\epsilon_\text{min}\right\}
\right],
\label{sda_delta}
\end{equation}
where $\alpha\Upsilon/\tau$ is given by
\begin{equation}
{\alpha\Upsilon\over\tau}={V_1B_1\over B_z^2}
{\Delta B_z\over\langle V_x\rangle}
={8r(r-1)\over(r+1)^3}.
\label{aconst}
\end{equation}
For $r=2.8$ in our shock, $e^{\alpha\Upsilon/\tau}\approx 2.09$ and $T_2=T_1e^{\alpha\Upsilon/\tau}=1.04$keV.

If we include the adiabatic heating in Eq.(\ref{esda}), then
the energy equation in the shock transition region becomes
\begin{equation}
{d\epsilon\over d  t}={\alpha(\epsilon-\epsilon_\text{min})\over\tau}
+{\epsilon\over\xi+t},\,\,\,(0<t<\Upsilon,\,\epsilon_1>\epsilon_\text{min}),
\label{ene_tot}
\end{equation}
where $\xi=\Upsilon/(r-1)$ and
the solution is given by
\begin{align}
\epsilon&=e^{\alpha t/\tau}\left(1+{t\over\xi}\right)
\left[\epsilon_1+\epsilon_\text{min}{\alpha\xi\over\tau}e^{\alpha\xi/\tau}
\left\{E_i\left(-{\alpha\xi\over\tau}\right)\right.\right. \nonumber \\ 
&\left.\left.-E_i\left(-{\alpha\over\tau}(t+\xi)\right)\right\}\right],
\,\,\,(0<t<\Upsilon,\,\epsilon_1>\epsilon_\text{min}),
\label{esol_tot}
\end{align}
where $E_i(s)\equiv \int_{-\infty}^{s}(e^q/q)dq$.

When the escape probability for an electron in the shock transition region is given by Eq.(\ref{delta}),
we calculate the energy distribution in the downstream using Eq.(\ref{fdown}) and (\ref{esol_tot}). We get a translated Maxwellian distribution with a temperature $T_2=T_1re^{\alpha\Upsilon/\tau}$,
\begin{align}
f_2(\epsilon_2)&={1\over T_1re^{\alpha\Upsilon/\tau}}
\text{exp}\left[-{1\over T_1re^{\alpha\Upsilon/\tau}}
\left\{\epsilon_2+\epsilon_\text{min}{r\alpha\xi\over\tau}e^{r\alpha\xi/\tau}
\right.\right. \nonumber \\
&\left.\left.\times\left(E_i\left(-{r\alpha\xi\over\tau}\right)
-E_i\left(-{\alpha\xi\over\tau}\right)\right)\right\}\right], 
\,\,\,(\epsilon_2>\epsilon_*),
\label{efdist_tot}
\end{align}
where $\epsilon_*=\epsilon(t=\Upsilon,\epsilon_1=\epsilon_\text{min})$ in Eq.(\ref{esol_tot}),
$\alpha\Upsilon/\tau$ is given by Eq.(\ref{aconst}), and
$\alpha\xi/\tau= 8r/(r+1)^3$.
For $r=2.8$ in our shock, $\epsilon_*=12.9$keV
and $T_2=T_1re^{\alpha\Upsilon/\tau}=2.9$keV.
The energy distribution in the downstream for 
$\epsilon_2<\epsilon_\text{min}$ is given by pure adiabatic heating,
\begin{equation}
f_2(\epsilon_2)={1\over rT_1}\text{exp}\left[-{\epsilon_2\over rT_1}\right], 
\,\,\,(\epsilon_2<\epsilon_\text{min}).
\label{efdist_adia}
\end{equation}

\begin{figure}
\includegraphics[scale=.32]{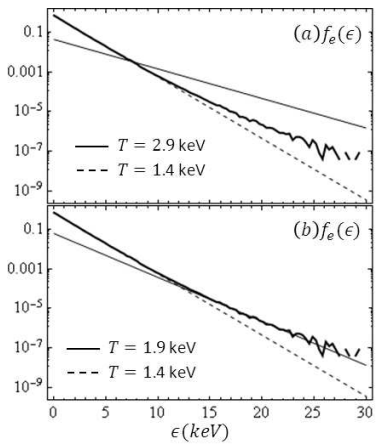}
\caption{Theoretical electron energy distribution $f_2(\epsilon)$
after SDA+adiabatic heating(solid line) and pure adiabatic heating(dashed line) with $r=2.8$ in (a). 
In (b), $r=2.1$ for SDA+adiabatic heating(solid line) and
$r=2.8$ for pure adiabatic heating(dashed line) are taken.
The simulation result(thick solid line)(Fig.\ref{edist}b) is plotted for comparison.} 
\label{sda_edist}
\end{figure} 
In Fig.\ref{sda_edist}, we plot Eq.(\ref{efdist_tot})(solid line), Eq.(\ref{efdist_adia})(dashed line) and the simulation result(thick solid line)[Fig.\ref{edist}b].
The temperature $T_2=2.9$keV for $\epsilon>\epsilon_*(=12.9$keV) from Eq.(\ref{efdist_tot}), however, is larger than the $T_2=1.9$keV from the simulation(Fig.\ref{edist}b).
The difference between the theory and the simulation results from the oscillatory shock structure in the simulation where the high energy electrons gain and lose energy via SDA by drifting downward($-\hat{y}$ axis) in the region of $\partial B/\partial x>0$ and by drifting upward($+\hat{y}$ axis) in $\partial B/\partial x<0$, respectively.
When an effective compression ratio, $r=2.1$, is applied to the electrons experiencing SDA, the corresponding temperature becomes
$T_2=T_1re^{\alpha\Upsilon/\tau}=1.9$ keV for $\epsilon>\epsilon_*(=8.6$keV), 
which is consistent with the simulation result.
In Fig.\ref{sda_edist}(b), we plot Eq.(\ref{efdist_tot})(solid line) with $r=2.1$, Eq.(\ref{efdist_adia})(dashed line) with $r=2.8$, and the simulation result(thick solid line).

\subsection{Effects on spectra when a  realistic mass ratio of protons and elections is used} 

Since our simulations have employed a reduce mass ratio of $m_i/m_e=30$, it is natural to wonder what the energy spectra would be for a real mass ratio $m_i/m_e=1836$ for fixed Mach number and plasma $\beta$. The energy spectrum depends on the shock structure, in particular the changes of $B_z$ and $E_x$ across the shock.  

The change in $B_z$ across the shock from $B_1$ to $rB_1$ and the shock electric potential jump $e\Delta\Phi$ in Eq.(\ref{potential_eq}) does not vary with the ion mass for fixed Mach number and plasma $\beta$. Thus, the same fraction of ions as computed for our reduced mass ratio will reflect at the shock front for the realistic mass ratio case. The MTSI still operates and
the growth rate is reduced by $\sqrt{30/1836}$ for fixed Mach number and plasma $\beta$. This implies that we need $\sqrt{1836/30}$ times longer time to generate the shock in the electron plasma frequency$(1/\omega_{pe})$ timescale.
In addition, the electron energy distribution via SDA in Eq.(\ref{efdist_tot}) 
as well as the minimum energy for the electrons to incur SDA in Eq.(\ref{emin}) does not vary with the ion mass.

In short, we expect the energy spectrum at least for electrons observed for our simulations with a reduced mass to be unchanged when a realistic mass is used.

\section{Conclusions}
\label{summary}

We simulated a purely perpendicular, low Mach number fast mode shock relevant to the termination shocks in solar flare magnetic reconnection outflows.
We employed a 2D particle-in-cell code with a moving wall boundary condition for a reduced ion/electron mass ratio $m_i/m_e=30$. The moving wall method generates a slowly propagating shock compared to the more standard fixed reflection boundary method, and allows for smaller box sizes and more efficient use of simulation time.

Both electrons and ions experience the shock drift acceleration(SDA) in our simulations. The transition energy point from pure adiabatic heating to SDA measured in the downstream rest frame is given by $\sim 5$ keV for the ions and
$\sim 10$ keV for the electrons.
These values are well modeled by our theoretical analysis of SDA. 
The negative $E_x$ field in the shock transition region 
requires the electrons to have high threshold energy for SDA. 
The ion energy distribution in the downstream shows a tri-Maxwellian distribution and the electron energy distribution in the downstream shows a bi-Maxwellian distribution.
We theoretically modeled the electron energy distribution via SDA with/without adiabatic heating.
If the probability for an electron to escape from the shock transition region
is given by a Dirac delta function, we have a bi-Maxwellian distributions which  agrees with the simulations.

The microphysical mechanism by which the collisionless shocks are sustained over the course of the simulation long after the initial shock formation stage is an important issue.
We have found that this is naturally explained by a modified two-stream instability due to the incoming and reflecting ions in the shock transition region--the unstable mode being therefore along the $k_x$ axis (shock normal direction) as seen from the spectral analysis of $E_x$ field. 
The maximum growth rate, $\gamma=4\times 10^{-4}\omega_{pe}$, from the modified two-stream instability is enough to excite the observed electric field fluctuation during the shock transit time($\sim 730/\omega_{pe}<1/\gamma$). 
We also found that the fluctuating field is responsible for the observed entropy generation throughout the downstream region

\subsection{Appendix - Moving Wall}
Particles will interact with the wall moving in the positive $x$ direction at some time between  $t_{n}$ and $t_{n+1}$  and not exactly at $t_{n+1}$.  We trace back both the particles position in space and the walls position in space, for each particle, and then calculate the correct reflection velocity used to determine the particles final and current locations.  The meeting location between wall and particle can be  found by determining the time increment $\Delta t_x = t_{n+1} -t_{meet}$  at which the particle collides with the wall using
\begin{equation}
 x_p-v_p\Delta t_x=x_{wall}-v_{wall}\Delta t_x,
 \end{equation}
 where $x_p$ is the particle location in $x$ at $t_{n+1}$, $v_p$
 is the particle's $x$ velocity at $t_{n+1}$, and $x_{wall}$ is given by $L_x + v_{wall}\Delta t$.  Once $\Delta t_x$ is determined the particle's position is advanced backward by an increment $v_p\Delta t_x$, the current deposited is removed and the proper reflection velocity is determined.
 
We determine the rebound velocities of the particles by Lorentz boosting into the frame co-moving with the wall.  In this frame, the $x$ component of the velocity is simply reversed and the $y$ and $z$ components of the velocity are  unchanged after a collision.  We then transform back to the simulation frame and determine the new velocities by equating the total momentum 4 vectors in each frame. 
The particles are then advanced by $v(\Delta t - \Delta t_x)$ and their momenta are calculated using the new Lorentz factor calculated from their rebound velocities.

\begin{acknowledgments}
We would like to thank Dr. Rui Yan and Wen Han for useful discussion and programming support. 
We thank the anonymous reviewer for valuable comments.
We also thank the OSIRIS consortium for the use of OSIRIS.
This work was supported by DOE under Grant DE-FG02-06ER54879 and Cooperate Agreement No. DE-FC52-08NA28302, by NSF under Grant PHY-0903797, and by NSFC under Grant No. 11129503. The research used resources of NERSC. 

\end{acknowledgments}


\begin{thebibliography}{}
\bibitem{priest02} E. R. Priest and T. G. Forbes, Astr. Astrophys. Rev. {\bf 10}, 313 (2002)

\bibitem{zharkova11} V. V. Zharkova, K. Arzner, A. O. Benz, \textit{et al}.
Space Sci. Rev. {\bf 159}, 357 (2011)

\bibitem{blackman1} E. G. Blackman and G. B. Field, Phys. Rev. Lett. {\bf 73}, 3097 (1994)

\bibitem{forbes88} T.G. Forbes, Solar Physics {\bf 117}, 97 (1988)

\bibitem{workman11} J. C. Workman, E. G. Blackman, and C. Ren, Phys. Plasmas {\bf 18}, 092902 (2011)

\bibitem{mann06} G. Mann, H. Aurass, and A. Warmuth, A\&A {\bf 454}, 969 (2006)

\bibitem{mann09} G. Mann, A. Warmuth, and H. Aurass, {\bf 494}, 669 (2009)

\bibitem{shibata95} K. Shibata, S. Masuda, M. Shimojo, H. Hara, T. Yokoyama, S. Tsuneta, T. Kosugi, and Y. Ogawara, ApJ {\bf 451}, L83 (1995)

\bibitem{decker86} R. B. Decker and L. Vlahos, ApJ {\bf 306}, 710 (1986)

\bibitem{amano07} T. Amano and M. Hoshino, ApJ {\bf 661}, 190 (2007)

\bibitem{kato08} T. N. Kato and H. Takabe, ApJ {\bf 681}, L93 (2008)

\bibitem{spitkovsky08} A. Spitkovsky, ApJ {\bf 673}, L39 (2008)

\bibitem{martins09} S. F. Martins, R. A. Fonseca, L. O. Silva, and W. B. Mori, ApJ {\bf 695}, L189 (2009)

\bibitem{sironi11} L. Sironi and A. Spitkovsky, ApJ {\bf 726}, 75 (2011)

\bibitem{amano09} T. Amano and M. Hoshino, Phys. Plasmas {\bf 16}, 102901 (2009). 

\bibitem{guo10} F. Guo and J. Giacalone, ApJ {\bf 715}, 406 (2010)

\bibitem{gargate12} L. Gargate and A. Spitkovsky, ApJ {\bf 744}, 67 (2012)

\bibitem{langdon88} B. Langdon, J. Arons, and C. Max, Phys. Rev. Lett. {\bf 61}, 7, (1988)

\bibitem{osiris02} R. A. Fonseca \textit{et al.}, Lect. Notes Comput. Sci. {\bf 2331}, 342 (2002).

\bibitem{tsuneta96} S. Tsuneta, ApJ {\bf 456}, 840 (1996)

\bibitem{tidman71}D. A. Tidman and N. A. Krall, Shock Waves in Collisionless Plasmas, p.14, p.10, and p.115, Wiley-Interscience, New York, (1971)

\bibitem{papadopoulos71} K. Papadopoulos, C. E. Wagner, and I. Haber, Phys. Rev. Lett. {\bf 27}, 982 (1971)

\bibitem{wagner71} C. E. Wagner, K. Papadopoulos, and I. Haber, Phys. Lett. A {\bf 35}, 440 (1971)

\bibitem{leroy82} M. M. Leroy, D. Winske, C. C. Goodrich, C. S. Wu, and
K. Papadopoulos, J. Geophy. Res. {\bf 87}, 5081 (1982)

\bibitem{hoshino01} M. Hoshino, Prog. Theo. Phys. Suppl. {\bf 143}, 149 (2001)

\bibitem{gary93} S. P. Gary, Theory of Space Plasma Microinstabilities, p.23, Cambridge Univ. Press, Cambridge(1993)

\bibitem{nakamura98} T. K. Nakamura and M. Hoshino, Phys. Plasmas {\bf 5}, 10 (1998)

\bibitem{jenkins70} M. A. Jenkins and J. F. Traub, Numer. Math. {\bf 14}, 252 (1970)

\bibitem{jenkins72} M. A. Jenkins and J. F. Traub, Algorithm 419: Comm. ACM {\bf 15}, 97 (1972)

\bibitem{bose09} S. K. Bose, Numeric computing in fortran, p.77, Alpha Science, Oxford, (2009)

\bibitem{parks12} G. K. Parks, E. Lee, M. McCarthy, M. Goldstein, S. Y. Fu, J. B. Cao, P. Canu, N. Lin, M. Wilber, I. Dandouras, H. R{\'e}me, and A. Fazakerley, Phys. Rev. Lett. {\bf 108}, 061102 (2012)

\bibitem{hoshino02} M. Hoshino and N. Shimada, ApJ {\bf 572}, 880 (2002)

\bibitem{dieckmann06} M. E. Dieckmann and P. K. Shukla, Plasma Phys. Control. Fusion {\bf 48}, 1515 (2006)

\bibitem{kato10} T. N. Kato and H. Takabe, ApJ {\bf 721}, 828 (2010)

\bibitem{webb83} G. M. Webb, W. I. Axford, and T. Terasawa, ApJ {\bf 270}, 537 (1983)

\bibitem{whipple86} E. C. Whipple, T. G. Northrop, and T. J. Birminham
JGR {\bf 91}, 4149 (1986) 

\bibitem{decker88} R. B. Decker, Space Science Reviews {\bf 48}, 195 (1988) 

\bibitem{begelman90} M. C. Begelman and J. G. Kirk, APJ {\bf 353}, 66 (1990) 

\bibitem{kirk94} J. G. Kirk, in Plasma Astrophysics, J. G. Kirk, D. B. Melrose, and E. R. Priest, Springer-Verlag(Berlin), p.225 (1994)

\bibitem{ball01} L. Ball and D. B. Melrose, Publ. Astron. Soc. {\bf 18}, 361 (2001)

\end{thebibliography}
\end{document}